\documentclass[pre,twocolumn,superscriptaddress]{revtex4-1}

\usepackage{amsmath}
\usepackage{mathtools} 
\usepackage{amssymb}
\usepackage{amsthm}
\usepackage{bm,upgreek} 
\usepackage{upgreek} 
\usepackage{xcolor}
\usepackage{graphicx}
\usepackage{epstopdf}
\usepackage[colorlinks,linkcolor=blue,anchorcolor=blue,citecolor=blue,urlcolor=blue]{hyperref}

\newcommand{\nn}{{\nonumber}}
\newcommand{\bea}{\begin{eqnarray}}
\newcommand{\eea}{\end{eqnarray}}
\newcommand{\ie}{\textit{i.e.{ }}}
\newcommand{\eg}{\textit{e.g.{ }}}

\newcommand{\av}[1]{\left\langle #1 \right\rangle}

\newcommand{\md}{\mathrm{d}}
\newcommand{\me}{\mathrm{e}}

\begin{document}

\title{Reducing autocorrelation time in determinant quantum Monte Carlo using
Wang-Landau algorithm: application to Holstein model}
\author{Meng Yao}
\affiliation{National Laboratory of Solid State Microstructures $\&$ School of Physics, Nanjing
University, Nanjing, 210093, China}
\author{Da Wang}\email{dawang@nju.edu.cn}
\affiliation{National Laboratory of Solid State Microstructures $\&$ School of Physics, Nanjing
University, Nanjing, 210093, China}
\affiliation{Collaborative Innovation Center of Advanced Microstructures, Nanjing University, Nanjing 210093, China}
\author{Qiang-Hua Wang}\email{qhwang@nju.edu.cn}
\affiliation{National Laboratory of Solid State Microstructures $\&$ School of Physics, Nanjing
University, Nanjing, 210093, China}
\affiliation{Collaborative Innovation Center of Advanced Microstructures, Nanjing University, Nanjing 210093, China}

\begin{abstract}
When performing a Monte Carlo calculation, the running time should in principle be much
longer than the autocorrelation time in order to get reliable results.
Among different lattice fermion models, the Holstein model is notorious for
its particularly long autocorrelation time.
In this work, we employ the Wang-Landau algorithm in the determinant quantum Monte Carlo to achieve the flat-histogram
sampling in the ``configuration weight space'',
which can greatly reduce the autocorrelation time by sacrificing some sampling efficiency.
The proposal is checked in the Holstein model on both square and
honeycomb lattices.
Based on such a Wang-Landau assisted determinant quantum Monte Carlo method, some
models with {long} autocorrelation times can now be simulated {possibly}.
\end{abstract}
\maketitle

When applying quantum Monte Carlo (MC) to correlated fermion systems, the negative sign problem is in general found to be nondeterministic polynomial hard \cite{troyer_computational_2005} as the sign average $\av{s}\propto\me^{-a\beta N}$ approaches zero exponentially versus the imaginary time $\beta$ (inverse of temperature) times particle number $N$, which inevitably demands exponential computational effort to get controlled data when approaching thermodynamic limit. Similarly, the same thing happens to the problems with exponential autocorrelation time even without sign problem. Moreover, if the autocorrelation time $\tau_a$ is sufficiently long, we cannot simply divide the whole Markov chain into several short ones (walkers) to reduce the statistical error unless the starting point of each walker can be chosen uniformly at random on the whole Markov chain\footnote{This is not in contradict with the ergodicity theorem which only states that the time average when $T\rightarrow\infty$ is equal to the ensemble (very huge phase space) average but not (only a few) walker average.}, which, however, is difficult to realize due to the nonuniform density of states (DOS) of the ``configuration weight space'' (corresponding to energy space in classical models).

When designing a MC algorithm, it is very important to make the autocorrelation time short enough. \cite{janke_monte_2008} To this end, roughly two classes of methods have been broadly used: {cluster algorithms} \cite{Swendsen_PRL_1987,Wolff_PRL_1989} and reweighting techniques\cite{Swendsen_PRL_1986,berg_multicanonical_1992,hukushima_exchange_1996}. Most of them work pretty well for discrete classical models such as Ising and Potts models, but for a quantum model, it is difficult to design a powerful method in general. Global update in principle can greatly reduce the autocorrelation time but always at the price of severely losing the algorithm efficiency or yielding very low acceptance rate. \cite{scalettar_ergodicity_1991} On the other hand, reweighting by another parameter set\cite{Swendsen_PRL_1986} or other replicas \cite{hukushima_exchange_1996} are often used to avoid the ergodicity breaking. In special, recently, reweighting by another trial model utilizing self-learning techniques was proposed to reduce the autocorrelation time efficiently. \cite{liu_self-learning_2017}

In this work, we focus on another kind of widely used reweighting technique, Wang-Landau (WL) algorithm\cite{wang_efficient_2001,wang_determining_2001}, and apply it to the determinant quantum Monte Carlo (DQMC) \cite{blankenbecler_monte_1981,assaad_world-line_2008}.
The WL algorithm provides a general and powerful strategy to obtain the flat-histogram simulation in the space of energy or some other given quantities for classical problems. For quantum models, the WL method has been successfully incorporated to the stochastic series expansion
\cite{troyer_flat_2003}, {diagrammatic Monte Carlo \cite{Diamantis_WL_Green_2013} and Renyi entropy calculations \cite{Inglik_WL_Renyi_2013}.}
In this work, we take another approach by directly generalizing the energy space in the classical case to the ``configuration weight space'' ($w$-space) in the quantum case.
Based on the WL method, the density of states $\rho(w)$ in the $w$-space can be obtained with high accuracy, which results in a flat-histogram sampling in the $w$-space by choosing $\rho^{-1}(w)$ as the MC sampling weight. Comparing to the usual importance (\eg Metrapolis or heat-bath) sampling, the autocorrelation time $\tau_a$ can be greatly reduced, although at the price of sacrificing some sampling efficiency. The advantage of such a WL assisted DQMC (WL-DQMC) algorithm is that some problems with super long autocorrelation times previously can now be reliably simulated by multi short walkers with high accuracy as a result of the reduced $\tau_a$.

As an application of our algorithm, we choose the {two-dimensional} half-filled Holstein model,
\begin{align}
H=&-t\sum_{\langle ij \rangle\sigma}(c_{i\sigma}^\dag c_{j\sigma}+{\rm h.c.}) + \Omega\sum_i b_i^\dag b_i \nn \\ &+\eta\sum_{i} (n_i-1)(b_i^\dag+b_i),
\end{align}
where $c_{i\sigma}$ and $b_i$ are electron and phonon field operators, $t$ is the hopping taken as $1$ below, $\Omega$ is the phonon frequency, and $\eta$ is the electron-phonon coupling. As usual, we define the dimensionless electron-phonon constant $\lambda=2\eta^2/\Omega W$ where $W$ is the band width of the free fermion model. The Holstein model can be simulated using the standard DQMC method {by sampling the displacement field directly \cite{johnston_determinant_2013}, which is free of sign problem} but notorious for its super long autocorrelation time. \cite{hohenadler_autocorrelations_2008}
However, recently, it received more and more attention for its underlying interesting properties such as charge density wave (CDW), superconductivity, and quantum critical point. \cite{Costa_PRL_2018,Karakuzu_HH_2018,chen_charge-density-wave_2019,zhang_charge_2019}
{Recently, there have also been some other QMC  approaches to the Holstein or Holstein-Hubbard model, including the interaction expanded continuous time QMC \cite{Assaad_ctqmc_2007,Weber_loop_2017,Weber_HH_2018,Hohenadler_HH_2019}, machine learning assisted DQMC \cite{chen_symmetry-enforced_2018,Li_ML_2019}, hybrid QMC \cite{Beyl_HMC_2018}, Langevin QMC \cite{Batrouni_Langevin_2019, CohenStead_Langevin_2020}.
In this work, we are intended to incorporate the WL algorithm to the standard DQMC and test it on the Holstein model.}

\begin{figure}
\begin{center}
\includegraphics[width=0.5\textwidth]{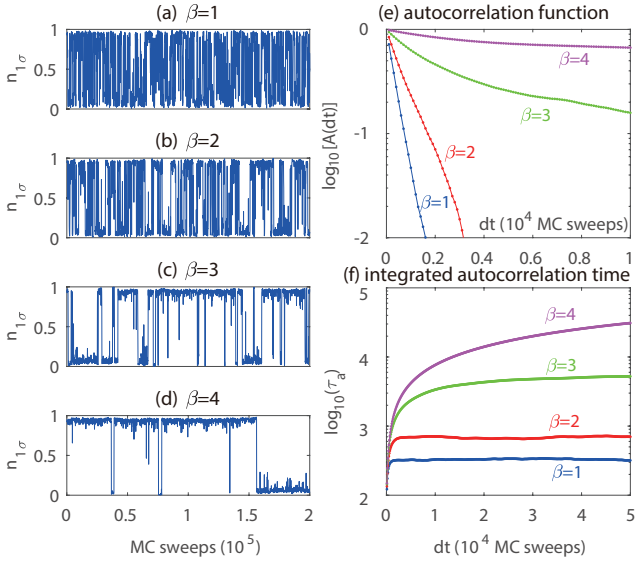}
\caption{DQMC simulattions of the Holstein model on the square lattice with $L=4$, $\Omega=1$ and $\lambda=0.5$. The discrete time slice $\Delta\tau=0.1$ in all simulations in this work. (a) to (d) plot the histogram of the particle number $n_{1\sigma}$. The normalized autocorrelation function $A(\md t)$ of $n_{1\sigma}$ is shown in (e), from which we extract the integrated autocorrelation times $\tau_a$ as defined in Eq.~\ref{eq:int_autocor} plotted in (f).
\label{fig:old}}
\end{center}
\end{figure}

We first show the ergodicity problem in the usual DQMC simulations on the square lattice. In Fig.~\ref{fig:old}(a)-(d), we plot the histogram of the particle number at one site $n_{1\sigma}$ from $\beta=1$ to $4$, respectively. Clearly, as $\beta$ increases, the change (from $\sim1$ to $\sim0$ or vice versa) of $n_{1\sigma}$ becomes more and more difficult, indicating the ergodicity problem.
It can be quantitatively characterized by the normalized autocorrelation function
\begin{align}
A(\md t)=\frac{\av{n_{1\sigma}(t)n_{1\sigma}(t+\md t)}}{\av{n_{1\sigma}^2}}.
\end{align}
In Fig.~\ref{fig:old}(e), $A(\md t)$ is plotted versus $\md t$, in which the exponential dependence of $A(\md t)\sim \me^{-\md t/\tau_a}$ is obtained. In this work, we follow the standard definition of the integrated autocorrelation time $\tau_a(\md t)$\cite{janke_monte_2008}, \ie
\begin{align}\label{eq:int_autocor}
\tau_a(\md t)=\frac12+\sum_{k=1}^{\md t}A(k),
\end{align}
which grows up as $\md t$ increases and finally saturates to a practical estimation of the autocorrelation time, as shown in Fig.~\ref{fig:old}(f). Roughly speaking, $\tau_a$ exponentially depends on $\beta$ reflecting the effect of the Boltzmann weight factor $\me^{-\beta\Delta}$ for tunneling  across a barrier $\Delta$. In these simulations, the global update by shifting a distance for each site with all time slices \cite{scalettar_ergodicity_1991,johnston_determinant_2013} has already been used, but the autocorrelation time is still too long. For the case of $\beta=4$, $\tau_a$ is at the order of $10^5$ MC sweeps, which demands the measurements more than $\sim10^7$ MC sweeps (much longer than $\tau_a$) for each walker in a serious MC simulation. Clearly, as $\beta$ increases, the MC simulations become more and more time consuming. Therefore, how to reduce the autocorrelation time is a very important problem.

In the DQMC method, by keeping physical boson fields or introducing auxiliary fields, the fermionic degrees of freedom can be integrated out, giving rise to the bosonic partition function $Z=\sum_c w(c)$ where $w(c)$ is in general can be negative or complex. Right now, we suppose real $w(c)>0$ for simplicity. The generalization to the cases with sign problem is straightforward. Defining the density of states (DOS) $\rho(w)=\sum_c\delta[w-w(c)]$, the partition function can be rewritten as $Z=\sum_w w\rho(w)$. The DOS $\rho(w)$ can be obtained using the standard WL algorithm.
After $\rho(w)$ is obtained with high accuracy, we obtain a flat-histogram sampling in the $w$-space by choosing $\rho^{-1}[w(c)]$ as the Markov chain weight. Then, a physical quantity $\av{A}$ becomes
\begin{align}
\av{A}&=\frac{\sum_c A(c)w(c)}{\sum_c w(c)}=\frac{\sum_c A(c) w(c) \rho[w(c)] \rho^{-1}[w(c)]}{\sum_c w(c) \rho[w(c)] \rho^{-1}[w(c)]} \nn\\
&\rightarrow\frac{\sum_i' A_i w_i \rho_i}{\sum_i' w_i\rho_i}=\frac{\av{Aw\rho}}{\av{w\rho}}
\label{eq:meas}
\end{align}
where $\sum'$ means summation on the Markov chain.
In practice, we record $\ln\rho$ and $\ln w$ (corresponding to the action) in computers to avoid numerical overflow. The algorithm is now described as follows:

Step-1. One performs a short DQMC run at first to get the range of $\ln w$ in the usual DQMC sampling.

Step-2. By choosing a larger WL-window $\ln w\in[W_{\rm min},W_{\rm max}]$ covering the obtained $\ln w$-range in step-1 to perform the standard WL algorithm: we accept a new configuration $c'$ with the probability $r=\min[\rho[w(c)]/\rho[w(c')],1]$. Since the weight ratio is already obtained in DQMC, this step causes no additional computational effort.

Step-3. For each sampled configuration with weight $w$, we update $\ln\rho(\ln w)=\ln\rho(\ln w)+\ln f$, where $f$ is initially chosen as $e$. After the histogram is ``almost'' flat, we set $f\rightarrow\sqrt{f}$ and repeat steps-2 and -3 until $f$ is sufficiently close to $1$, say $f<1+10^{-6}$. Then, the desired DOS $\ln\rho(\ln w)$ is obtained.

Step-4. We stop updating $\ln\rho$ and begin to do measurements using $\rho^{-1}$ as the sampling weight. Notice that according to Eq.~\ref{eq:meas}, we need accumulate $A_iw_i\rho_i$ and $w_i\rho_i$ simultaneously to get the physical quantity $\av{A}=\av{Aw\rho}/\av{w\rho}$.

\begin{figure}
\begin{center}
\includegraphics[width=0.52\textwidth]{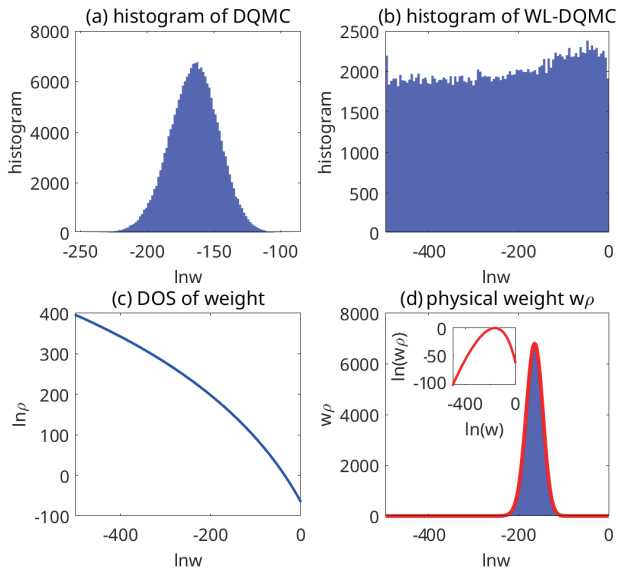}
\caption{Histograms of the configuration weight $\ln w$ are plotted in (a) for DQMC and (b) for WL-DQMC. (c) DOS $\ln\rho$ obtained by WL-DQMC versus $\ln w$. (d) By multiplying $w$ and $\rho$, we obtain the physical weight which is the same as in (a) shown as the shaded area.
\label{fig:wl}}
\end{center}
\end{figure}

As an example to clarify the algorithm, let us focus on the square lattice with $L=4$, $\lambda=0.5$ and $\beta=4$. {With this parameter setup, we will show that the autocorrelation time $\tau_a$ is neither too short (standard DQMC applies) nor too long (WL-DQMC also fails) but in a reasonable range: of order} $10^5$ MC sweeps in DQMC even with the global update.
At first, we perform a usual DQMC run. The histogram is plotted in Fig.~\ref{fig:wl}(a), which tells us the physically important region is almost $\ln w\in[-250,-100]$. Then by choosing a larger WL-window $[-500,-50]$, we perform the WL sampling to obtain the DOS $\ln \rho$ versus $\ln w$ as shown in Fig.~\ref{fig:wl}(c). Interestingly, the DOS is not bounded in the WL-window and continues to grow up as $\ln w$ decreases. This behavior is not changed for moderately larger WL-windows. In fact, for the Holstein model, there should be no lower bound of $\ln w$ since the phonon displacement can be arbitrarily large. With the obtained $\rho^{-1}$ as the sampling weight, we do obtain the desired flat histogram as shown in Fig.~\ref{fig:wl}(b). By multiplying $w$ and $\rho$, we can reconstruct the ``physical weight'' which is in good agreement with the usual importance sampling, as shown in Fig.~\ref{fig:wl}(d).

\begin{figure}
\begin{center}
\includegraphics[width=0.4\textwidth]{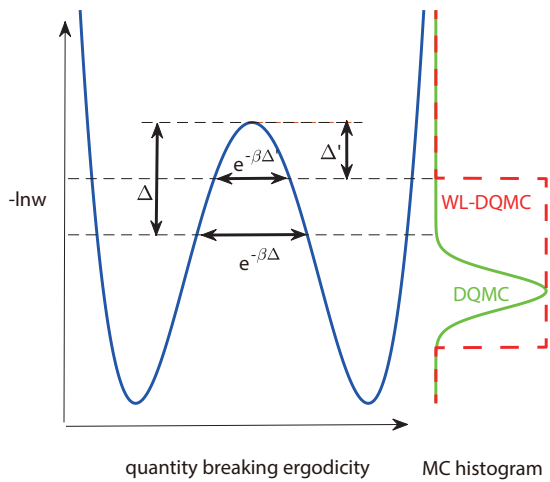}
\caption{A schematic plot showing how WL-DQMC increases the tunneling probability and thus reduces the autocorrelation time for exponential barrier problems. $-\ln w$ (action) takes the role of free energy in the classical case. The MC histogram of WL-DQMC is plotted with solid green (dashed red) curves on the right.
\label{fig:scheme}}
\end{center}
\end{figure}

At first glance, the WL-DQMC enlarges the sampling region in the $w$-space and thus performs many ``useless'' measurements, which inevitably decreases the sampling efficiency. However, the lower-weight configurations (corresponding to higher-energy configurations classically) can reduce the tunneling barrier between different local minimums in the phase space. As depicted in the schematic diagram Fig.~\ref{fig:scheme}, the tunneling probability $\me^{-\beta\Delta}$ in the usual DQMC can be greatly enlarged to $\me^{-\beta\Delta'}$ by the WL-DQMC sampling. Therefore, the autocorrelation time is greatly reduced. In fact, it can be expected that if the WL-window is large enough, the tunneling gap $\Delta'$ vanishes and the transition probability only depends on $\beta$ polynomially.
Comparing with the usual reweighting technique with a higher temperature\cite{Swendsen_PRL_1986} or another group of parameters, the WL-DQMC has its advantage that the most important configurations are fully kept, while the sampling efficiency (roughly the area of DQMC over WL-DQMC in the histogram in Fig.~\ref{fig:scheme}) only decreases inversely proportional to the WL-window size.

\begin{figure}
\begin{center}
\includegraphics[width=0.48\textwidth]{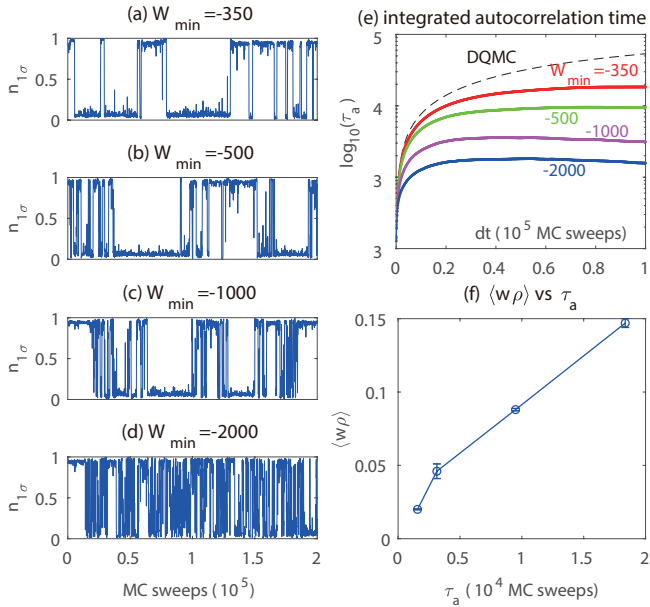}
\caption{Reducing autocorrelation time using the WL-DQMC with different WL-windows $[W_{\rm min},-50]$. (a)-(d) plot the histograms of $n_{1\sigma}$ with different $W_{\rm min}$. The integrated autocorrelation time $\tau_a$ is shown in (e) together with the DQMC result for comparison. In (f), we plot the average $\av{w\rho}$ versus $\tau_a$, which clearly shows the decrease of $\tau_a$ simultaneously accompanies the decrease of $\av{w\rho}$.
\label{fig:wldqmc}}
\end{center}
\end{figure}

The above picture is fully consistent with our numerical results. As shown in Fig.~\ref{fig:wldqmc}(a)-(d), by decreasing the lower bound of the WL-window $W_{\rm min}$ while keeping $W_{\rm max}=-50$, we indeed observed that the change of $n_{1\sigma}$ occurs more and more frequently. Correspondingly, the integrated autocorrelation time $\tau_a$ drops significantly up to two orders (for $W_{\rm min}=-2000$ to $\sim10^3$ MC sweeps) relative to the usual DQMC, as shown in Fig.~\ref{fig:wldqmc}(e). Of course, the price of the decrease of $\tau_a$ is the drop of $\av{w\rho}$, as shown in Fig.~\ref{fig:wldqmc}(f). However, as long as $\av{w\rho}$ is not too small, say $\av{w\rho}>10^{-3}$, we can still get reliable data by the means of statistical (multi-walker) average.
{The total efficiency of the WL-DQMC algorithm is determined by these two factors: $\tau_a$ and $\av{w\rho}$. But as $\tau_a$ can be reduced exponentially while $\av{w\rho}$ only drops algebraically, their combination is still anticipated to be efficient.}

\begin{figure}
\begin{center}
\includegraphics[width=0.48\textwidth]{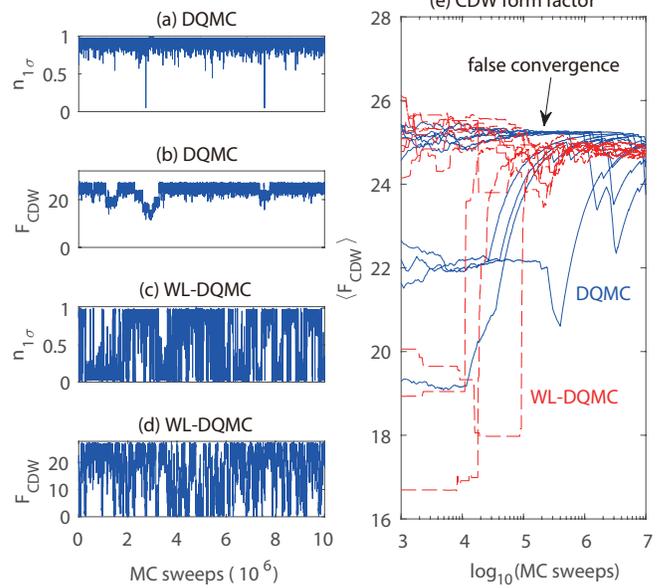}
\caption{The WL-DQMC simulations on the honeycomb lattice with $L=4$, $\lambda=2/3$, and $\beta=8$. (a) and (b) plot the histograms of $n_{1\sigma}$ and $F_{\rm CDW}$ obtained in the usual DQMC simulations with the global update included. Similarly, (c) and (d) are obtained in our WL-DQMC algorithm with WL-window $[W_{\rm min}, W_{\rm max}]=[-2000,-500]$. In (f), the MC averaged CDW form factor $F_{\rm CDW}$ are plotted versus the MC sweeps in ten walkers for DQMC (solid blue lines) and WL-DQMC (dashed red lines), respectively.
\label{fig:honeycomb}}
\end{center}
\end{figure}

As another application, we turn to the half-filled honeycomb lattice which was extensively studied very recently. \cite{chen_charge-density-wave_2019,zhang_charge_2019} We choose one typical group of parameters: $L=4$, $\lambda=2/3$, and $\beta=8$ for comparing with Ref.~\onlinecite{zhang_charge_2019}. As shown in Fig.~\ref{fig:honeycomb}(a), the particle number of one site $n_{1\sigma}$ is pinned at $\sim1$ within $10^7$ MC sweeps. Similarly, the CDW form factor $F_{\rm CDW}=\frac{1}{2L^2}\sum_{i}(-1)^{i+j}\av{n_i n_j}$ is also pinned at values from $24$ to $26$ as shown in Fig.~\ref{fig:honeycomb}(b). This is a clear feature of the ergodicity breaking and indicates that the autocorrelation time $\tau_a$ should be at least larger than $10^7$ MC sweeps, causing great difficulty in MC simulations. As an approximation, the researchers performed simulations with multi short runs to do average. \cite{zhang_charge_2019} However, the approximation error is difficult to estimate. By trying different WL-windows, we found $[W_{\rm min},W_{\rm max}]=[-2000,-500]$ can reduce the autocorreltion times of $n_{1\sigma}$ and $F_{\rm CDW}$ to $\sim1.5\times10^5$ and $\sim3\times10^4$ MC sweeps (not shown), respectively. This now enabled us to perform more reliable MC simulations. In Fig.~\ref{fig:honeycomb}(e), we plot the results of $\av{F_{\rm CDW}}$ versus MC sweeps in ten walkers using WL-DQMC and DQMC, respectively. As anticipated by the shorter $\tau_a$, we did observe that the ten WL-DQMC walkers almost converge at order of $10^6$ MC sweeps. Their difference comes mainly from statistical error. In contrast, the ten DQMC walkers give quite inconsistent results up to $10^7$ MC sweeps as a result of the super long autocorrelation time. It should be mentioned that if we take the ``false convergence'' [six of ten walkers, indicated by the arrow in Fig.~\ref{fig:honeycomb}(e)] as an approximation, the error is found to be small. Therefore, we conclude that in this specific model the approximated DQMC simulations with running time much shorter than autocorrelation time \cite{zhang_charge_2019} are still qualitatively correct.

In summary, we implemented the WL algorithm in the DQMC framework to realize the flat-histogram sampling in the configuration weight space. The advantage of such a WL-DQMC method is to greatly reduce the autocorrelation time for exponential barrier problems. Its feasibility is checked in the Holstein model on both square and honeycomb lattices. With the WL-DQMC algorithm, some problems with {long} autocorrelation times previously can now be simulated {possibly}.
Of course, our present work is a direct application of the WL algorithm in the DQMC by replacing the energy space to configuration weight space. Clearly, such an idea can also be applied to other quantum Monte Carlo algorithms as long as the flat-histogram is suitably chosen. As a future direction, we can achieve the flat-histogram in the space of some other physical quantities, \eg order parameters which may be more efficient to recover the ergodicity.

D.W. thanks Zhichao Zhou, Yuxi Zhang, Tianxing Ma, {Chuang Chen and Zi-Yang Meng} for helpful discussions.
This work is supported by the NSFC under Grant Nos. 11874205 and 11574134.
The numerical simulations were performed in
High-Performance Computing Center of Collaborative
Innovation Center of Advanced Microstructures, Nanjing University.

\bibliography{wl-dqmc}

\end{document}